\documentstyle[11pt,pictex]{article}

\def\labelmark{}

\def\void{}
\newenvironment{formula}[1]{\def\labelname{#1}
\ifx\void\labelname\def\junk{\begin{displaymath}}
\else\def\junk{\begin{equation}\label{\labelname}}\fi\junk}%
{\ifx\void\labelname\def\junk{\end{displaymath}}
\else\def\junk{\end{equation}}\fi\junk\labelmark\def\labelname{}}

{\ifx\void\labelname\def\junk{\end{array}\end{displaymath}}
\else\def\junk{\end{array}\right.\end{equation}}
\fi\junk\labelmark\def\labelname{}\def\junk{}
}

\newcommand{\beq}{\begin{formula}}
\newcommand{\eeq}{\end{formula}}

\newcommand{\beqa}{\begin{eqnarray}}
\newcommand{\eeqa}{\end{eqnarray}}
\newcommand{\eq}[1]{(\ref{#1})}
\newcommand{\nn}{\nonumber} 
\newcommand{\alim}[1]{\stackrel{{#1}\rightarrow\infty}
{\longrightarrow}}

\newcommand{\define}{\stackrel{\mbox{def.}}{=}}
\newcommand{\tmu}{{\bar\mu}}
\newcommand{\ra}{\rightarrow}
\newcommand{\der}{\partial}
\newcommand{\eps}{\epsilon}
\catcode`\*=11  
\def\slashsym#1#2{\mathpalette{\sl*sh{#1}}{#2}}
\def\sl*sh#1#2#3{\ooalign{\setbox0=\hbox{$#2\not$}
                          $\hfil#2\mkern-24mu\mkern#1mu
                           \raise.15\ht0\box0\hfil$\cr
                          $#2#3$}}
\catcode`\*=12

\def\kslash{{\slashsym7k}}

\def\pslash{{\slashsym8p}}

\def\epslash{{\slashsym7\epsilon}}
\newcommand{\NP}[1]{ {\it Nucl.~Phys.} {\bf #1}}
\newcommand{\PL}[1]{ {\it Phys.~Lett.} {\bf #1}}

\newcommand{\PR}[1]{ {\it Phys.~Rev.} {\bf #1}}
\newcommand{\PRL}[1]{ {\it Phys.~Rev.~Lett.} {\bf #1}}

\newcommand{\AP}[1]{ {\it Ann.~Phys.} {\bf #1}}
\newcommand{\JMP}[1]{ {\it J. Math.~Phys.} {\bf #1}}
\newcommand{\WS}[1]{ (World Scientific Publishing,#1)}
\begin{document}

\begin{titlepage}
\setcounter{page}{0}
\renewcommand{\thefootnote}{\fnsymbol{footnote}}

\begin{flushright}
\mbox{\phantom{draft v7}}\\
HD-THEP-98-41\\
hep-th/9809053\\
(revised)
\end{flushright}

\vspace{5 mm}
\begin{center}
{\Large  Integral representations of thermodynamic 1PI Green 
functions in the world-line formalism}

\vspace{10 mm}

{\bf Haru-Tada Sato$^{}$
\footnote{E-mail: sato@thphys.uni-heidelberg.de }}
\vspace{5mm}

\vspace{10mm}

{\it $^{}$ Institut f{\"u}r Theoretische Physik\\
Universit{\"a}t Heidelberg\\
Philosophenweg 16, D-69120 Heidelberg, Germany}
\end{center}

\vspace{10mm}
\begin{abstract}
The issue discussed is a thermodynamic version of the 
Bern-Kosower master amplitude formula, which contains all 
necessary one-loop Feynman diagrams. It is demonstrated how the 
master amplitude at finite values of temperature and chemical 
potential can be formulated within the framework of the 
world-line formalism. In particular we present an elegant 
method how to introduce a chemical potential for a loop in the 
master formula. Various useful integral formulae for the master 
amplitude are then obtained. The non-analytic property of the 
master formula is also derived in the zero temperature limit 
with the value of chemical potential kept finite. 
\end{abstract}

\vfill

\begin{flushleft}
\end{flushleft}
\end{titlepage}
\setcounter{footnote}{0}
\renewcommand{\thefootnote}{\arabic{footnote}}
\renewcommand{\theequation}{\thesection.\arabic{equation}}
\section*{1. Introduction}\label{sec1}
\setcounter{section}{1}
\setcounter{equation}{0}
\indent

Based on the field theory limit of string theory (with infinite 
string tension limit), a very elegant method was invented several 
years ago: the Bern-Kosower (BK) rules to obtain one-loop gluon 
scattering amplitudes in a compact form ~\cite{BK} 
(see also~\cite{sum1}). {}~For example, five gluon 
scatterings are efficiently calculated by using these 
rules~\cite{five}, and various field theory limits have been 
studied along the line of the BK formalism: perturbative 
gravity~\cite{gra}, super Yang-Mills theory~\cite{sYM}, 
bosonic string theory approach~\cite{renorm}, and multi-loop 
generalizations \cite{kaj,RS,paolo}. 

The most important and conspicuous point in the BK formalism is 
that all Feynman diagrams are included in a single master 
integrand. In this formalism, we hence do not compute the loop 
integrals and the Dirac traces of respective Feynman diagrams. 
These facts can also be achieved by another approach, called the 
world-line formalism, which reformulates Feynman (field theory) 
amplitudes similar to string theory amplitudes; i.e. the field 
theory amplitudes can generally be obtained as a path integral 
average of vertex operators~\cite{SSphi}-\cite{YM2}.

In fact, there have been established many examples: $\phi^3$ 
theory \cite{SSphi,myphi}, QED \cite{SSqed,sumino}, axial vector 
and pseudo-scalar couplings \cite{pseudo,MS}, Yang-Mills 
theory~\cite{YM1,YM2} (see also \cite{book,tset}), and more 
references can be found in \cite{sum2}. Both world-line and 
string theory methods help each other, and are useful to get 
an insight for solving mutually related problems in each own 
way; for example, this viewpoint has been very useful for 
specifying pinching limits and corners of string moduli in the 
multi-loop analysis~\cite{RS}. It is very interesting that 
these two methods, which entirely differ from the conventional 
Feynman diagram technique, improve the computational efficiency 
for obtaining Feynman amplitudes. 
 
However, compared to these developments, their thermodynamic 
versions have not been studied very much~\cite{KST}-\cite{sho} 
from the viewpoint of general formulation for constructing a 
master amplitude formula. In particular, there has been no 
general and convenient method how to introduce a chemical 
potential for a loop in the world-line formalism. Before 
stepping in an unexplored calculation by using the formalism, 
it is important to establish a definite and universal foundation 
in the first place. In this paper, therefore as a basic step on 
the thermodynamic generalization of world-line field theory, we 
present a universal and fundamental prescription of one-loop 
$N$-point amplitudes at finite values of temperature and 
chemical potential without help of any standard calculation. 
We shall study the amplitude of a particular form (the master 
formula), where the loop integration and the Matsubara summation 
are {\it a priori} finished and Feynman's parameter integrals are 
only left --- laying emphasis on the point that we never mean 
the Feynman integrals as a single Feynman diagram, but as a sum 
of {\it all diagrams}. It is certainly nontrivial to introduce a 
chemical potential (as well as a temperature) with keeping this 
advantageous point pertaining to the master formula intact. 

We address the following points in the matter of the 
thermodynamic generalization (at finite values of the 
temperature $\beta^{-1}$ and the chemical potential $\mu$ for 
a loop). First, we present a formulation of the thermodynamic 
amplitudes along the same line as the non-thermodynamic 
world-line formalism. In particular, the way of introducing the 
chemical potential is a nontrivial problem. Since we do not 
introduce any idea of the continuous or discrete momentum 
integration/summation, we have to find an alternative to the 
shift of internal discrete momenta:
\beq{matsu}
\omega_n\quad\ra\quad\omega_n+i\mu\ .
\eeq
This situation may be understood in the following way: In the 
standard method, the inverse temperature $\beta$ is 
introduced by summing up all topological different $S^1$ paths 
along the zeroth component (imaginary time) direction. 
On the other hand in the world-line formalism, this procedure 
modifies the path integral of a corresponding periodic world-line 
field $x^0(\sigma)$; $0\leq\sigma\leq1$, into a summation of 
the path integrals with $x^0(\sigma)$ shifted by 
\beq{step1}
x^0(\sigma)\quad\ra\quad x^0(\sigma)+n\beta\sigma\ .
\eeq 
Although 
the radius $\beta$ of $S^1$ and the world-line circumference 
(the unity) have nothing to do with each other, the 
shift~\eq{step1} involves the world-line coordinate $\sigma$ 
as well. In this sense, the way of prescribing the temperature 
and hence of the chemical potential become different from the 
standard Matsubara formalism. To introduce the chemical potential, 
one may of course transform a world-line amplitude formula to 
the Feynman-Matsubara form, and then resume the world-line form 
after some efforts to apply the shift~\eq{matsu}. However, such 
a calculation does not utilize the merits of the master formula 
at all, and there should be a more direct and transparent method 
to introduce $\mu$ within the world-line formalism (without 
tracing back and referring any internal loop calculation).

To this end, we propose a new rule, instead of \eq{matsu}, 
to introduce the internal chemical potential. It is simply 
done by applying the new shift procedure
\beq{step2}
{\bar\omega}\quad\ra\quad{\bar\omega}+i\mu\ ,
\eeq
where ${\bar\omega}$ is an average of the zeroth components of 
external continuous/discrete momenta $k^0_j$; $j=1,2,\cdots N$, 
with the summation weights $\sigma_j$ (the local coordinates 
of the external legs on the closed loop). The prescription 
\eq{step2} is neither conceivable nor explicable from the 
standard method \eq{matsu}, because $\omega_n$ is the internal 
momentum while ${\bar\omega}$ concerns the external one. 
The parameter ${\bar\omega}$ will easily be identified in due 
course, if we adopt a statistical parameter to discern between 
boson and fermion loops when summing up the $S^1$ paths: This 
parameter dependence should vanish at the $\beta\ra\infty$ limit 
as expected in the non-thermodynamic world-line formalism. 

Another non-triviality in this formalism is how to derive a 
non-analytic property at $\beta=\infty$ with finite $\mu$ 
from the master amplitude formula. Since the non-analytic 
property can be derived from the $\beta=\infty$ limits of pure 
thermodynamic parts, we first have to separate a pure 
thermodynamic part ${\tilde\Gamma}_N^{\beta\mu}$ from a full 
thermodynamic amplitude $\Gamma_N^{\beta\mu}$. If the master 
integrand of $\Gamma^{\beta\mu}_N$ is composed only of the Jacobi 
$\Theta$-function, the story is simple as expected. 
However, in a more complicated case like a photon scattering, 
the master integrand is not such a simple form but a product of 
a $\beta$- and $\mu$-dependent operator ${\cal V}_{\beta\mu}$ 
and the Jacobi $\Theta$-function part ${\cal K}_{\beta\mu}$. 
The pure thermodynamic parts of these quantities, 
${\tilde{\cal V}}_{\beta\mu}$ and ${\tilde{\cal K}}_{\beta\mu}$, 
can easily be separated from their original full quantities, 
but we show that the pure thermodynamic 
part ${\tilde\Gamma}_N^{\beta\mu}$ is non-trivially given by 
${\cal V}_{\beta\mu}\times{\tilde{\cal K}}_{\beta\mu}$ 
against naive expectation. (The simple arithmetical splittings 
indicate one more contribution, however we shall show that 
it vanishes). Separating the ${\tilde\Gamma}_N^{\beta\mu}$ 
in this way, we then analyze the non-analytic property of 
the master amplitude at zero temperature. 

This paper is organized as follows. 
In Section~\ref{sec2}, we explain our notations, definitions and 
the general structure of master amplitudes at zero temperature. 
In Section~\ref{sec3}, we derive a set of general formulae 
for a master amplitude at finite $\beta$ and $\mu$ parts 
by parts: path integral normalization and (scalar) kinematical 
factor in Section~\ref{sec3a}, and reduced kinematical 
factor in Section~\ref{sec3b}. In Section~\ref{sec3c}, 
we show a general master formula for the purely thermodynamic 
part of the full amplitude $\Gamma_N^{\beta\mu}$, and also prove 
the above statement, i.e., ${\tilde\Gamma}_N^{\beta\mu}
\sim {\cal V}_{\beta\mu}\times{\tilde K}_{\beta\mu}$. 
The replacement rule \eq{step2} is verified in Appendices A-C 
from the viewpoints of both Feynman rule's calculation and the 
world-line formalism. In Sections~\ref{sec4} and \ref{sec5}, 
for arbitrary $N$, we derive various integral formulae 
by analyzing two kinds of $\Theta$-function representations, 
and check their consistency. Several explicit results are also 
presented up to $N=5$. In Section~\ref{sec6}, we derive
the non-analytic property of the master formula in the zero 
temperature limit with the chemical potential kept finite. 
Section~\ref{sec7} contains summary and conclusions.

\section{Notations and definitions}\label{sec2}
\setcounter{section}{2}
\setcounter{equation}{0}
\indent

{}~First, we summarize the general structure of the one-loop 
$N$-point master amplitudes~\footnote{We assume that a particle 
change does not occur while circulating along the loop.} 
in the non-thermodynamic world-line formalism. 
We refer the reader to Refs.~\cite{SSqed,YM1,sum2} 
for more details. The master amplitude of general form 
(for $N$ external momenta $k^\mu_j$; $j=1,2,\cdots,N$; 
$\mu=0,1,\cdots,D-1$) is written in terms of the closed path 
integrals of bosonic $x^\mu(\sigma)$ and fermionic 
$\psi^\mu(\sigma)$ world-line fields as follows: 
\beq{master}
\Gamma_N={1\over2}\int_0^\infty{ds\over s}
\oint {\cal D}x^\mu(\sigma){\cal D}\psi^\mu(\sigma)\,
e^{-\int_0^1 {\cal L}(\sigma)d\sigma}\prod_{j=1}^N V_j
\eeq
with the world-line Lagrangian and the vertex operators
\beqa
{\cal L}(\sigma)&=&{1\over4s}\Bigl(
{\der x^\mu(\sigma)\over\der\sigma} \Bigr)^2
+{1\over2}\psi^\mu(\sigma){\der\over\der\sigma}\psi_\mu(\sigma)
+sm^2\ ,\\
V_j &=& s\int_0^1d\sigma \,v_j[x(\sigma),\psi(\sigma)]\,
e^{ik_j\cdot x(\sigma)} \ ; \qquad j=1,2,\cdots,N\ ,
\eeqa
where $k_j\cdot x$ stands for the Lorentz contraction, and we 
often omit the Lorentz indices as long as obvious. 
The zero mode integral of the bosonic path integral should be 
excluded~\cite{book}. The explicit form of $v_j$ depends on 
what particle is inserted in the loop as an external leg; for 
example, $v_j=1$ for $\phi^3$ theory, and is in Eq.\eq{photon} 
for the photon vertex case. Note that our world-line coordinate 
$\sigma$ is dimensionless, and is related to the standard 
notation~\cite{sum2} by scaling $\tau=s\sigma$. 
{}~For the path integral average of a general quantity $F$, 
\beq{*} 
<F(x,\psi)>\equiv{\cal N}^{-1}\oint{\cal D}x{\cal D}\psi
 e^{-\int_0^1 {\cal L}(\sigma)d\sigma}\,F(x,\psi)\ ,
\eeq
one may use the Wick contractions with~\footnote{
The Euclidean metric is given by $g^{\mu\nu}=-\delta^{\mu\nu}$.} 
\beqa
&& <x^\mu(\sigma_1)\,x^\nu(\sigma_2)>=
   -sg^{\mu\nu}G(\sigma_1,\sigma_2)\ ,\\
&&<\psi^\mu(\sigma_1)\psi^\nu(\sigma_2)>=
{1\over2}g^{\mu\nu}\mbox{sign}(\sigma_1-\sigma_2)\ ,
\eeqa
where ${\cal N}$ is the path integral normalization 
\beq{Norm}
{\cal N}=\oint{\cal D}x{\cal D}\psi
 e^{-\int_0^1 {\cal L}(\sigma)d\sigma}\,
= e^{-sm^2}(4\pi s)^{-D/2}\ ,
\eeq
and $G$ the bosonic world-line correlator 
\beq{G12}
G(\sigma_i,\sigma_j)=|\sigma_i-\sigma_j|-(\sigma_i-\sigma_j)^2 
\equiv G_{ij}\ .
\eeq

Performing the path integrals (or Wick contractions), we arrive at 
the following master amplitude formula: 
\beq{gammaN}
\Gamma_N = c\int_0^1 d\sigma_1 \int_0^1 d\sigma_2
\cdots\int_0^1 d\sigma_N\int_0^\infty ds\, s^{N-1}
{\cal V}\times {\cal K}\ ,
\eeq
where the constant $c$ depends on a theory~\footnote{
It is also related to an over-counting factor of 
$\sigma$-integration regions \cite{norm}.} 
(particle circulating in a loop); For example,
\beq{*}
c=\left\{ 
\begin{array}{ll}
{1\over2} &\quad\mbox{for}\quad\mbox{a (neutral) scalar loop}\\
-{1\over2}\mbox{tr}[1] &\quad \mbox{for}\quad
\mbox{a fermion loop,} 
\end{array}\right. 
\eeq
where the $\mbox{tr}[1]$ expresses the trace of a unit matrix in 
the $D$-dimensional gamma matrix space. {}~For a gluon loop, 
$c$ is no longer a simple constant~\cite{YM1,YM2}. 
The quantity ${\cal K}$, which we shall call the 
{\it kinematical factor} (multiplied by 
the normalization ${\cal N}$), 
is defined by the world-line path integral average 
of $N$ scalar vertex operators~\cite{SSphi,YM1} 
\beq{norm}
{\cal K}\equiv{\cal N}<\prod_{j=1}^N e^{ik_jx(\sigma_j)}>
=\oint{\cal D}x^\mu{\cal D}\psi^\mu\,(\prod_{j=1}^N 
e^{ik_j\cdot x(\sigma_j)})
 e^{-\int_0^1 {\cal L}(\sigma)d\sigma}\ .
\eeq
The ${\cal V}$ stands for an effective vertex function, 
which is obtained by
\beq{defV}
{\cal V}={<\prod_{j=1}^Nv_j \exp[ik_j\cdot x(\sigma_j)]>
\over <\prod_{j=1}^N \exp[ik_j\cdot x(\sigma_j)]>} \ .
\eeq
We shall refer to this quantity as the 
{\it vertex structure function/operator}, 
which corresponds to the quantities called the 
{\it reduced kinematical factor}~\cite{BK} or 
the generating kinematical factor~\cite{YM1}. 
At this stage, the ${\cal V}$ is still a function of $s$ and 
$\sigma_j$ ($j=1,2,\cdots,N$), however it will be generalized 
to an operator in the thermodynamic case. 

If the vertex structure function can be expanded in the 
form
\beq{Vexp}
{\cal V}=\sum_{l\in {\bf Z}} a_l\, s^{l-N}\,\exp[-sb_l]\ ,
\eeq
where the coefficients $a_l$ and $b_l$ may not depend on $s$, 
but on $\sigma_j$, the amplitude \eq{gammaN} is obtained as the 
sum of the 'partial' amplitudes 
\beq{Al}
{\cal A}_l \define \int_0^\infty dss^{l-1}{\cal K}
=\int_0^\infty ds s^{l-1}{\cal N}<\prod_{j=1}^N 
e^{i k_j\cdot x(\sigma_j)}> \ ,
\eeq
with shifting $m^2$ to $m^2+b_l$.

In the case of finite $\beta$, we assume the zeroth components 
of the external momenta $k^\mu_j$ 
to be the bosonic Matsubara frequencies~\footnote{
We do not consider the 
fermionic external states, since the bosonic external states 
are only well-formulated in the world-line formalism. However 
one may formally generalize to the fermionic states.} 
\beq{omegan}
k^0_j=\omega_{k_j}={2\pi\over\beta}n_j\ ,
\eeq
as well as 
\beq{*}
k_j\cdot x\equiv \omega_{k_j}x^0-\vec{k}_j\cdot\vec{x}\ .
\eeq
We use the notations for the counterparts of Eqs.
\eq{gammaN}, \eq{norm}, \eq{Norm}, \eq{defV}, \eq{Al}
\beq{*}
\Gamma_N^\beta,  \quad {\cal K}_\beta,\quad
{\cal N}_\beta,\quad {\cal V}_\beta, \quad 
{\cal A}_l^\beta \ .
\eeq
In the case of finite $\mu$ and $\beta$, we denote 
\beq{*}
\Gamma_N^{\beta\mu},  \quad {\cal K}_{\beta\mu},\quad
{\cal N}_{\beta\mu},\quad {\cal V}_{\beta\mu}, \quad 
{\cal A}_l^{\beta\mu} \ .
\eeq
We explain how to define these quantities in the next section. 

\section{The thermodynamic generalization}\label{sec3}
\setcounter{section}{3}
\setcounter{equation}{0}
\subsection{The kinematical factor ${\cal K}_{\beta\mu}$ and 
the normalization ${\cal N}_{\beta\mu}$}\label{sec3a}
\indent

We take a two-step procedure to introduce the temperature and the 
chemical potential. As the first step, we consider the $\mu=0$ 
case. Basically we follow the same method as discussed in Refs.
\cite{MR}-\cite{sho}, and we start with the following slightly 
general definition:
\beq{path}
\Gamma_N^\beta \define \sum_{n=-\infty}^\infty 
e^{2n\pi i\over\epsilon}\, 
\Gamma_N\,\Bigr|_{x^0(\sigma)\ra x^0(\sigma)+n\beta\sigma}\ ,
\eeq
where $\epsilon$ is a constant related to the statistics of the 
loop. {}~For example, the $\epsilon=2$ case corresponds to a 
fermion loop, and the $\epsilon=\infty$ case to a bosonic loop. 
Without specifying the value of $\epsilon$, we deal with both 
cases simultaneously (formally fractional statistics as well). 
The $\Gamma_N$ on the right-hand side (r.h.s.) of the above 
formula denotes the path integral representation \eq{master}, 
and the replacement \eq{step1} should be applied to the $x^0$ 
fields for all variables $\sigma_j$. In a nutshell, we have 
only to replace the bosonic path integral in the following way:
\beq{boint}
\oint{\cal D}x^\mu\quad\ra\quad 
\oint_\beta{\cal D}x^\mu\equiv \sum_{n=-\infty}^\infty 
e^{2n\pi i\over\epsilon}\, 
\oint_{x^0(\sigma)\ra x^0(\sigma)+n\beta\sigma} 
\hskip-60pt{\cal D}x^\mu\hskip50pt .
\eeq

{}~For simplicity, let us consider the ${\cal V}=1$ case, or the 
'partial' amplitude ${\cal A}_N$ (the $l=N$ term in \eq{Vexp}).
In this case, we have only to generalize the 
kinematical factor ${\cal K}$ to the finite temperature version 
${\cal K}_\beta$;
\beqa
{\cal K}_\beta 
&\define& \sum_{n=-\infty}^\infty e^{2n\pi i\over\epsilon}
{\cal N}<\prod_{j=1}^N e^{ik_j\cdot x(\sigma_j)} > 
\Bigr|_{x^0(\sigma)\ra x^0(\sigma)+n\beta\sigma}\  \\
&=& (4\pi s)^{-D\over2}
e^{-sM^2}\Theta_3({1\over\epsilon}+{\beta{\bar\omega}\over2\pi}, 
i{\beta^2\over4\pi s})\ , \label{ampb}
\eeqa
where the definition of $\Theta_3(v,\tau)$ is 
\beq{def}
\Theta_3(v,\tau)=\sum_{n=-\infty}^\infty
      e^{n^2\tau\pi i}e^{2nv\pi i}\ ,\label{def}
\eeq
and we have introduced the following two $s$-independent quantities:
\beqa
M^2 &=& m^2-\sum_{i<j,=1}^Nk_i\cdot k_j G_{ij}\ ,\\
{\bar\omega} &=& \sum_{j=1}^N \sigma_j\omega_{k_j}\ . \label{omebar}
\eeqa
Here the two-point correlator (world-line Green function) 
$G_{ij}$, given by Eq.\eq{G12}, looks different from the one used 
in Ref.\cite{MR}, however the value of $M^2$ does not differ 
under the condition of momentum conservation regarding the 
external legs. The sign of $M^2$ seems not always to be positive 
in spite that $0\leq G_{ij} \leq1/2$. We then assume $M^2\geq0$ 
in the following discussion, choosing the off-shell symmetric 
point (satisfying the momentum conservation constraint) 
\beq{kk}
k_ik_j=\delta_{ij}k^2+(\delta_{ij}-1){1\over N-1}k^2\ .
\eeq
{}~For the ${\cal V}\not=1$ case, 
we have to apply Eq.\eq{boint} to the ${\cal V}$ and ${\cal K}$ 
parts at the same time, however the things are rather 
straightforward. We refer the reader to Appendix C for 
an example. 

The second step is to introduce the chemical potential for the 
internal loop. One may do it with applying the Jacobi 
transformation to the quantity ${\cal K}_\beta$; i.e., rewriting 
Eq.\eq{ampb} in such a way to revive the discrete summation 
over internal Matsubara frequencies $\omega_n$, one may perform 
the replacement \eq{matsu} (see \cite{HS} for more details). 
However, this is a roundabout way, since one has to make the 
Matsubara summation re-appear in spite of dealing with the 
world-line formulation, where the integration and summation of 
the loop are already performed. Instead, we propose a 
much simpler and direct alternative method to steer clear of 
this problem. It can be promptly done by the replacement~\eq{step2}, 
\beq{shift}
{\bar\omega} \quad\ra\quad \Omega \equiv {\bar\omega} +i\mu \ .
\eeq
This shift looks similar to the one \eq{matsu}, however we stress 
that our shifting parameter is not the internal frequency 
but an average of external ones (cf. Eq.\eq{omebar}). 
In this sense, this prescription is nontrivial. 
{}~For a rigorous reader, we put a justification of this 
replacement in Appendix A (the case of ${\cal V}=1$), and 
also refer to Appendix B for more complicated case 
(the ${\cal V}\not=1$ case). After all, the general path 
integral average for finite $\mu$ can be obtained in a couple 
of simple replacements: 
\beq{*}
\oint{\cal D}x^\mu \quad\ra\quad \oint_\beta{\cal D}x^\mu
\quad\ra\quad \oint_{\beta\mu}{\cal D}x^\mu
\equiv\oint_\beta{\cal D}x^\mu\,\Bigr|_{{\bar\omega}\ra\Omega}\ .
\eeq
Note that the fermion path integral does not change by any means. 

Let us write down the thermodynamic kinematical factor 
${\cal K}_{\beta\mu}$ and the normalization factor 
${\cal N}_{\beta\mu}$. Applying the above replacement 
to Eq.\eq{ampb}, we thereby yield the 
desired thermodynamic kinematical factor for finite $\beta$ 
and $\mu$: 
\beqa
{\cal K}_{\beta\mu} 
&\define& \mbox{Eq.}\eq{ampb}\,\Bigr|_{{\bar\omega}\ra\Omega} \\
&=& (4\pi s)^{-D\over2}
e^{-sM^2}\Theta_3(i{\beta\tmu\over2\pi},i{\beta^2\over4\pi s})\ ,
\label{firstrep}
\eeqa
where we have introduced the shorthand notation $\tmu$ by reason 
of analogy to the vacuum amplitude of a scalar loop
\beq{tmu}
\tmu \equiv \mu-i{\bar\omega} - {2\pi i\over\epsilon\beta} 
=-i(\Omega+{2\pi\over\epsilon\beta})\ .
\eeq
We refer to the representation \eq{firstrep} as the 
{\it first representation}. This representation with $\tmu=0$ 
($\mu={\bar\omega}=0$, $\eps=\infty$) is studied in Ref.~\cite{MR}. 
We also obtain another expression (which we shall call the 
{\it second representation}) through the Jacobi transformation,
\beq{secondrep}
{\cal K}_{\beta\mu} = {1\over\beta}(4\pi s)^{1-D\over2}
e^{-s(M^2-\tmu^2)}
\Theta_3({2s\tmu\over\beta},i{4\pi s\over\beta^2})\ .
\eeq
Note that for a fermion loop ($\epsilon=2$), the above first and 
the second $\Theta_3$ representations become the $\Theta_4$ 
and $\Theta_2$ representations respectively. 

We can extract the thermodynamic quantity ${\cal N}_{\beta\mu}$ 
corresponding to the path integral normalization~\eq{Norm} from 
the kinematical factor ${\cal K}_{\beta\mu}$. If we rewrite 
Eq.\eq{firstrep} as
\beqa
{\cal K}_{\beta\mu} &=& e^{-sm^2}(4\pi s)^{-D/2}\Theta_3(
i{\beta\tmu\over2\pi}, i{\beta^2\over4\pi s})
\exp[\,s\sum_{i<j}^N k_i\cdot k_jG_{ij}\,] \nn\\
&=&{1\over\beta}(4\pi s)^{1-D\over2}e^{-s(m^2-\tmu^2)}
\Theta_3({2s\tmu\over\beta},{4\pi is\over\beta^2})
<\prod_{j=1}^N e^{ik_j\cdot x(\sigma_j)}>\ ,
\eeqa
the non-correlator part can be regarded as an 
overall normalization to the $N$-point correlator of zero 
temperature type (the zeroth components $k^0_j$ are formally 
regarded as continuous variables here). 
Paraphrasing this fact in analogy to Eq.\eq{norm}
\beq{*}
{\cal K}_{\beta\mu} \equiv {\cal N}_{\beta\mu}
<\prod_{j=1}^N e^{ik_j\cdot x(\sigma_j)}>
= \oint_{\beta\mu}{\cal D}x{\cal D}\psi 
e^{-\int_0^1 {\cal L}(\sigma) d\sigma}
\prod_{j=1}^N e^{ik_j\cdot x(\sigma_j)} \ ,
\eeq
the thermodynamic version of the path integral normalization 
(for finite $\beta$ and $\mu$) is found to be 
\beq{normb}
{\cal N}_{\beta\mu} = 
{1\over\beta}(4\pi s)^{1-D\over2}e^{-s(m^2-\tmu^2)}
\Theta_3({2s\tmu\over\beta},{4\pi is\over\beta^2})\ .
\eeq
This is exactly the same normalization as assumed in 
Ref.\cite{KST} (taking a mass term inclusion into account). 

\subsection{The vertex structure operator 
${\cal V}_{\beta\mu}$}\label{sec3b}
\indent

In this subsection, we discuss the part of vertex structure 
(${\cal V}\not=1$). Because of diversities of explicit forms 
of ${\cal V}$, there is no concrete formula such as 
Eq.\eq{firstrep}. However, we can derive a general property of 
the pure thermodynamic part ${\tilde{\cal V}}_{\beta\mu}$. To 
explain this, we classify ${\cal V}$ into two categories by the 
criterion whether or not ${\cal V}$ contains local world-line 
variables $\sigma_j$. 

{}~First, let us start with the first category, the 
$\sigma$-independent ${\cal V}$ case. Apparently, 
${\cal V}=1$ is the case. A nontrivial example of this category 
is the $\pi^0\ra2\gamma$ decay ($\epsilon=2$, $D=4$) without 
background field \cite{HS}:
\beq{*}
c{\cal V}=-tr[1]m\lambda e^2 \epsilon_{\mu\nu\rho\sigma}
\epsilon_1^\mu\epsilon_2^\nu k_1^\rho k_2^\sigma\ ,
\eeq  
where $\lambda$ is the pseudo-scalar coupling, $m$ and $e$ the 
(space-time) fermion mass and the QED coupling constant. 
Another nontrivial example is the effective potential of a 
fermion loop ($\epsilon=2$, $N=0$) in a constant magnetic field 
in $D$ dimensions \cite{KST}. In this 
case, ${\cal V}$ depends on the integration variable $s$, 
\beq{*}
c{\cal V}= -{1\over2}\mbox{tr}[1]\,sB\mbox{coth}(sB)\ ,
\eeq
and this is the case of the expansion \eq{Vexp}, if we use 
\beq{*}
\mbox{coth}(sB)= 1+2\sum_{n=1}^\infty e^{-2nsB}\ .
\eeq
With the shift $M^2\ra M^2+2lB$, this case is essentially 
described by the 'partial' amplitude ${\cal A}_1$ defined by 
Eq.\eq{Al}. Now, at finite values of $\beta$ and $\mu$, we 
can, in principle, express any thermodynamic vertex structure 
function ${\cal V}_{\beta\mu}$ as 
\beq{decV}
{\cal V}_{\beta\mu}={\cal V}+{\tilde{\cal V}}_{\beta\mu}\ ,
\eeq
where ${\tilde{\cal V}}_{\beta\mu}$ denotes the purely 
thermodynamic part. However, in this category, as can be seen 
from the above examples, we simply have 
\beq{*}
{\cal V}_{\beta\mu}={\cal V}\ ,\qquad
{\tilde{\cal V}}_{\beta\mu}=0 \quad
\qquad(\mbox{for $\sigma$-independent ${\cal V}$})\ ,
\eeq 
because the ${\cal V}$'s of this category do not 
contain the local world-line variables $\sigma_j$, strictly 
speaking the $G_{ij}$ which are generated by the bosonic field 
correlation from $v_j$ --- Recall that the $\beta$-dependence 
only appears through the shift $x^0\ra x^0+n\beta\sigma$ in 
the formula \eq{path}. The $\beta$-dependence can not be 
created from the quantity which does not contain a 
$x$ field correlation. 

Second, let us consider the other category, the $\sigma$-dependent 
${\cal V}$ case. In this category, we obtain a nonzero structure 
function ${\tilde{\cal V}}_{\beta\mu}$. We here consider the 
photon polarization case by way of example. In this case, 
the ${\cal V}$ is given by Eq.\eq{Vsphoto} at zero 
temperature~\cite{YM1}: 
\beq{*}
{\cal V}=\eps_1\cdot\eps_2{1\over s}{\ddot G}_{12}
+\eps_1\cdot k_2\eps_2\cdot k_1({\dot G}_{12})^2 
+\eps_1\cdot\eps_2 k_1\cdot k_2-\eps_1\cdot k_2\eps_2\cdot k_1\ ,
\eeq
where dots on $G_{12}$ means the first and the second derivatives 
w.r.t. the first argument of $G_{12}$. The thermodynamic 
generalization can be done by applying the formulae \eq{path} 
and \eq{shift}. {}~For ease of presentation, we 
put the computational details in Appendix C, and a comparison 
with the Feynman diagram technique is also in Appendix B. 
{}For finite $\beta$, we derive Eq.\eq{Vsbphoto}:
\beq{*}
{\cal V}_\beta = {\cal V} 
-\eps_0^1\eps_0^2({1\over s}{\der\over\der{\bar\omega}})^2 
-{1\over s}{\der\over\der{\bar\omega}}(\eps_0^1\eps_2\cdot k_1-
\eps_0^2\eps_1\cdot k_2){\dot G}_{12}\ ,
\eeq
and then shifting ${\bar\omega}\ra\Omega$, we acquire 
the operator part given by Eq.\eq{Vsbmphoto}: 
\beq{*}
{\tilde{\cal V}}_{\beta\mu}=
-\eps_0^1\eps_0^2({1\over s}{\der\over\der\Omega})^2 
-{1\over s}{\der\over\der\Omega}(
\eps_0^1\eps_2\cdot k_1-\eps_0^2\eps_1\cdot k_2){\dot G}_{12}\ .
\eeq
Note that the origin of $\der/\der\Omega$ is the Wick contractions 
of the bosonic world-line fields furnished in the $v_j$ parts 
of photon vertex operators (see Eqs.\eq{photon} and \eq{detail}), 
and it always happens if a vertex operator comprises a bosonic 
field in such a way. 

We therefore conclude that if ${\cal V}$ includes $G_{ij}$ or 
its derivatives, ${\tilde{\cal V}}_{\beta\mu}$ gives rise to 
a differential polynomial in $\der/\der\Omega$, and otherwise 
${\tilde{\cal V}}_{\beta\mu}=0$. An important result followed 
from this fact is 
\beq{vazero}
{\tilde{\cal V}}_{\beta\mu}\times {\cal K}=0
\qquad\qquad(\mbox{for all ${\cal V}$})\ .
\eeq
This is a model-independent result, and we shall make use of this 
relation in order to decouple the pure thermodynamic 
part ${\tilde\Gamma}_N^{\beta\mu}$ from the total $N$-point 
amplitude $\Gamma_N^{\beta\mu}$.

\subsection{The master formulae ${\tilde\Gamma}_N^{\beta\mu}$ 
and ${\tilde{\cal A}}_l^{\beta\mu}$}\label{sec3c}
\indent

Gathering the formulae obtained in the above sections, we compose 
the purely thermodynamic master amplitude 
${\tilde\Gamma}_N^{\beta\mu}$, and define the thermodynamic 
'partial' amplitudes ${\cal A}_l^{\beta\mu}$. 
Applying the following formula to the first representation 
~\eq{firstrep}
\beq{decomp}
\Theta_3(v,\tau)=1+2\sum_{n=1}^\infty e^{n^2\tau\pi i} 
\mbox{cos}(2n\pi v)\ ,
\eeq
we separate the pure thermodynamic part 
${\tilde{\cal K}}_{\beta\mu}$ from ${\cal K}_{\beta\mu}$ as 
\beq{decK}
{\cal K}_{\beta\mu}={\cal K}+{\tilde{\cal K}}_{\beta\mu}\ ,
\eeq
where 
\beq{pureKt}
{\tilde{\cal K}}_{\beta\mu} =
2 (4\pi s)^{-D\over2}e^{-sM^2}
\sum_{n=1}^\infty e^{-{n^2\beta^2\over4s}}\mbox{cosh}
(n\beta\tmu) \ , 
\eeq
or for the second representation~\eq{secondrep}
\beq{hoshi}
{\tilde{\cal K}}_{\beta\mu} =
{2\over\beta} (4\pi s)^{1-D\over2}e^{-s(M^2-\tmu^2)}
\sum_{n=1}^\infty e^{-s({2n\pi\over\beta})^2}\mbox{cos}
({4n\pi s\tmu \over \beta}) \ . 
\eeq
{}~For a given ${\cal V}_{\beta\mu}$, the thermodynamic master 
amplitude $\Gamma_N^{\beta\mu}$ is calculated as 
\beq{GBM}
\Gamma_N^{\beta\mu} \define 
c\int_0^1d\sigma_1\int_0^1d\sigma_2\cdots
\int_0^1d\sigma_N\int_0^\infty ds s^{N-1}
{\cal V}_{\beta\mu}\times{\cal K}_{\beta\mu} \ .
\eeq 
Using the decompositions \eq{decV} and \eq{decK} with the 
general formula \eq{vazero}, we find 
\beq{*}
\Gamma_N^{\beta\mu}=\Gamma_N+{\tilde\Gamma}_N^{\beta\mu}\ ,
\eeq
where
\beq{GMBt} 
{\tilde\Gamma}_N^{\beta\mu} \define
c\int_0^1d\sigma_1\int_0^1d\sigma_2\cdots
\int_0^1d\sigma_N\int_0^\infty ds s^{N-1}
{\cal V}_{\beta\mu}\times{\tilde{\cal K}}_{\beta\mu} \ .
\eeq 
Therefore we have only to separate the pure thermodynamic part 
of ${\cal K}_{\beta\mu}$ in order to obtain the purely 
thermodynamic part ${\tilde\Gamma}_N^{\beta\mu}$: no separation 
in the ${\cal V}_{\beta\mu}$ part at all (Note 
the difference between Eqs.\eq{GBM} and \eq{GMBt}). 

As can be inferred from the examples in Section~\ref{sec3b}, 
a general expansion form of ${\cal V}_{\beta\mu}$ is 
\beq{Vbmexp}
{\cal V}_{\beta\mu}=\sum_{l,n\in {\bf Z}} 
a_{ln}\, s^{l-N}\, {\der^n\over\der\Omega^n}\,
\exp[-sb_l]\ .
\eeq
where the coefficients $a_{ln}$ and $b_l$ may not depend on $s$, 
but on $\sigma_j$ ($j=1,2,\cdots,N$). In analogy to Eq.\eq{Al}, 
the relevant `partial' amplitudes for Eqs.\eq{GBM} and \eq{GMBt} 
are defined as
\beq{AN}
{\cal A}_l^{\beta\mu} \define \int_0^\infty ds\, s^{l-1}
{\cal K}_{\beta\mu}\,\Bigr|_{m^2\ra m^2+b_l}\ ,
\eeq
\beq{AN2}
{\tilde{\cal A}}_l^{\beta\mu} \define \int_0^\infty ds\, s^{l-1}
{\tilde{\cal K}}_{\beta\mu}\,\Bigr|_{m^2\ra m^2+b_l}\ ,
\eeq
and of course the following relation holds:
\beq{*}
{\cal A}_l^{\beta\mu}={\cal A}_l + {\tilde{\cal A}}_l^{\beta\mu}\ .
\eeq

In the following sections, we present various integral 
representations for the pure thermodynamic parts 
${\tilde A}_l^{\beta\mu}$ of the $N$-point `partial' 
amplitudes without specifying any values 
of $N$, $D$ and $\epsilon$. These parts are the essential 
quantities to analyze the zero temperature limits with 
revealing the non-analyticity on $\mu$. 

\section{The integral formulae from the first 
representation}\label{sec4}
\setcounter{section}{4}
\setcounter{equation}{0}
\indent

In this section, we derive various integral formulae for 
${\tilde{\cal A}}_l^{\beta\mu}$ based on the first 
representation~\eq{firstrep}. This representation is examined 
in the special case $l=1$ (with $N=0$ and $\eps=2$), and 
actually $(4\pi)^{D/2}{\tilde{\cal A}}_1^{\beta\mu}$ is the 
function ${\cal O}_\beta$ analyzed in Ref.\cite{KST}: 
\beq{*}
{\cal O}_\beta(m)= 4\sum_{n=1}^\infty (-1)^n \mbox{cosh}(n\beta\mu)
({n\beta\over2m})^{1-D/2}K_{D/2-1}(n\beta m)\ .
\eeq
We want to generalize this function to more generic $l$ and 
$\eps$ cases, and it is convenient to mimic the computational 
technique of Ref.\cite{KST}, introducing the parallel notation 
\beq{OA}
{\cal O}_\beta^{(k)}(M)
\define (4\pi)^{D\over2}
{\tilde{\cal A}}_l^{\beta\mu}\ ;\qquad\quad k \equiv 2l+1-D\ .
\eeq
Hereafter, for simplicity we set
\beq{*}
       b_l=0 \ ,  
\eeq
and for later convenience, we also define
\beq{*}
            d \equiv 3-k = D+2-2l \ . 
\eeq

{}~From Eqs.\eq{pureKt} and \eq{AN2},the pure thermodynamic 
part ${\tilde{\cal A}}_l^{\beta\mu}$ is now given by 
\beq{*}
{\cal O}_\beta^{(k)}(M) =
2(4\pi)^{D\over2}
\int_0^\infty ds s^{l-1}(4\pi s)^{-D\over2}e^{-sM^2}
\sum_{n=1}^\infty e^{-{n^2\beta^2\over4s}}\mbox{cosh}
(n\beta\tmu) \ . 
\eeq
Performing the $s$-integration, we obtain
\beq{Obetak}
{\cal O}_\beta^{(k)}(M)=4\sum_{n=1}^\infty\mbox{cosh}(n\beta\tmu)
({n\beta\over2M})^{1-d/2}K_{d/2-1}(n\beta M)\ ,
\eeq
where $K_\nu(z)$ is the modified Bessel function of second kind. 
This is a generalized version (for arbitrary $\eps$ and $l$) 
of the above function ${\cal O}_\beta$. The aim is to obtain 
integral representations with performing the summation in 
Eq.\eq{Obetak} (for generic $k$ or $d$). To this end, 
we first apply the following formula to Eq.\eq{Obetak}:
\beq{Knuz}
K_\nu(z)=
{\sqrt\pi(z/2)^\nu\over\Gamma(\nu+{1\over2})}
\int_1^\infty e^{-zt}(t^2-1)^{\nu-{1\over2}}dt\ ; 
\quad\mbox{Re}\,\nu>-{1\over2}\ , \mbox{Re}\,z>0\ ,
\eeq
and note that the $k<2$ $(d>1)$ and $k\geq2$ $(d\leq1)$  
cases involve different calculations. 
Here we put a few remarks. {}~For the convergence of the summation 
over $n$, we have to assume 
\beq{*}
         \mu < M 
\eeq
in order to satisfy 
the condition 
\beq{*}
|e^{-\beta(Mt\pm\tmu)}|=e^{-\beta(Mt\pm\mu)}<1\ ;\qquad t>1\ .
\eeq 
The $k=1,0,-1$ ($d=2,3,4$) cases are calculated in Ref.\cite{KST}, 
and the $k=3$ ($d=0$) case corresponds to Ref.\cite{HS}, although 
they did not discuss this representation (Note that their 
argument belongs to our second representation). 
  
(i) In the first case, $k<2$ ($d>1$), the condition on $\nu$ in 
the formula \eq{Knuz} is satisfied as it is; i.e., 
$\nu={d\over2}-1>-{1\over2}$, and the calculation is almost 
parallel to Ref.\cite{KST}. We then just write down the result  
\beq{basicform}
{\cal O}_\beta^{(k)}(M)={-2\sqrt\pi\over\Gamma({d-1\over2})}\Bigl[
\int_0^\infty {(u^2+2Mu)^{d-3\over2}\over
1-e^{\beta(u+M+\tmu)}}du+(\tmu\ra-\tmu)\,\Bigr]\ ,
\eeq
where note that the minus sign of the denominator 
differs from the previous formula (Eq.(3.13) in Ref.\cite{KST}). 
For later convenience, 
let us derive another formula from this result. Performing 
the change of variable $u=M(\sqrt{p^2+1}-1)$, and applying 
the formula $\Gamma(z)\Gamma(1-z)=\pi/\mbox{sin}\pi z$ 
with $z=(d-1)/2$, we derive 
\beq{o1stcase1}
{\cal O}_\beta^{(k)}(M)=
{2\over\sqrt\pi}\Gamma({k\over2})M^{1-k}\mbox{sin}({2-k\over2}\pi)
\int_0^\infty {p^{1-k}(p^2+1)^{-1/2}\over e^{\beta X_+(p)}-1}dp 
+(\tmu\ra-\tmu)\ ,
\eeq
where we have introduced the compact notation 
\beq{*}
X_\pm(p)=M\sqrt{p^2+1}\pm\tmu \ .
\eeq

(ii) In the second case, $2\leq k$ ($d\leq1$), the calculation 
is similar to the case (i), excepting the point that we apply 
the formula \eq{Knuz} for $\nu=1-{d\over2}$ instead of 
$\nu={d\over2}-1$ owing to the relation $K_\nu(z)=K_{-\nu}(z)$. 
Then the summation over $n$ in Eq.\eq{Obetak} leads to the Lerch 
transcendental function: 
\beqa
{\cal O}_\beta^{(k)}(M)&=&
\alpha\int_1^\infty dt(t^2-1)^{1-d\over2}
e^{-\beta(Mt+\tmu)}\Phi(e^{-\beta(Mt+\tmu)},d-2,1)
+(\tmu\ra-\tmu)   \nn \\
&=&{\alpha\Gamma(3-d)\over2\pi i}\int_1^\infty 
dt(t^2-1)^{1-d\over2}
\int_\infty^{(0+)}
{(-z)^{d-3}dz\over1-e^{z+\beta(Mt+\tmu)}}
+(\tmu\ra-\tmu)\ ,\label{Lerch}
\eeqa
where just for conciseness we have defined the coefficient
\beq{*}
\alpha ={\sqrt{4\pi}\over\Gamma({3-d\over2})}
({\beta\over2})^{2-d}=
{\sqrt{4\pi}\over\Gamma({k\over2})}({\beta\over2})^{k-1}\ .
\eeq
The $z$-integrand \eq{Lerch} possesses poles at $z=0$ and 
$z=-\beta Mt\pm\beta\tmu+2n\pi i$ \, ($n\in {\bf Z}$). Since 
$-\beta(Mt\pm\mu)$ is a negative value, there is no pole in the 
positive region on the real axis other than $z=0$ as long as 
$\beta{\bar\omega}+{2\pi\over\epsilon}$ is not an integer. 
With the change of variable $t=\sqrt{p^2+1}$, and a replacement 
of one derivative $\der/\der z$ by $\der/\der p$, we have
\beq{sfint}
{\cal O}_\beta^{(k)}(M)
=\alpha{(-1)^{-k}\Gamma(k)\over M\beta(k-1)!}
{\partial^{k-2}\over\partial z^{k-2}}
\int_0^\infty p^{k-2} {\partial\over\partial p}
\Bigl({1\over1-e^{z+\beta X_+}}\Bigr)dp\Bigr|_{z=0}
+(\tmu\ra-\tmu)\ .
\eeq
Because of $k\geq2$, the surface terms from a partial integral 
vanish in Eq.\eq{sfint}, and finally we find 
\beq{o1stcase2}
{\cal O}_\beta^{(k)}(M)=2\Gamma({k+1\over2})
{(-1)^{1-k}\beta^{k-2}(k-2)\over M(k-1)!}
{\partial^{k-2}\over\partial z^{k-2}}
\int_0^\infty {p^{k-3}\over1-e^{z+\beta X_+}}dp\Bigr|_{z=0}
+(\tmu\ra-\tmu) \ .
\eeq
More explicitly, we present the results for $k=2,3,4$ as follows:
\beqa
{\cal O}_\beta^{(2)}(M)&=&-{\sqrt\pi\over M} \Bigl[
{1\over1-e^{\beta(M+\tmu)}} +(\tmu\ra-\tmu)\Bigr]\ ,\label{obk2}\\
{\cal O}_\beta^{(3)}(M)&=&{\beta\over M}\int_0^\infty 
{e^{\beta X_+}\over(1-e^{\beta X_+})^2}dp 
+(\tmu\ra-\tmu)\ ,\label{obk3}\\
{\cal O}_\beta^{(4)}(M)&=&-{\sqrt\pi\over2M^3}\Bigl[
{1\over1-e^{\beta(M+\tmu)}}-{\beta Me^{\beta(M+\tmu)}\over
(1-e^{\beta(M+\tmu)})^2} +(\tmu\ra-\tmu)\Bigr]\ .\label{obk4}
\eeqa
{}~For even values of $k$, one may directly perform the 
summation \eq{Obetak} without using the integral 
representation \eq{Knuz}; e.g., using $K_{1/2}(z)=\sqrt{\pi/2z}
e^{-z}$ and $K_{3/2}(z)=\sqrt{\pi/2z}(1+z^{-1})e^{-z}$ for 
$k=2$ and 4. We will see in the next section that the results 
\eq{o1stcase1} and \eq{obk2}-\eq{obk4} can be reproduced from 
the second representation as well.

\section{The integral formulae from the 
second representation}\label{sec5}
\setcounter{section}{5}
\setcounter{equation}{0}
\indent

In this section, based on the second 
representation~\eq{secondrep} case, we derive some more formulae 
on ${\cal O}_\beta^{(k)}(M)$, with re-deriving the results 
of the previous section. {}~From 
Eqs.\eq{hoshi}, \eq{AN2}, and \eq{OA}, we have the following 
form of the pure thermodynamic part to start with:
\beq{pureO}
{\cal O}_\beta^{(k)}(M)={(4\pi)^{1/2}\over\beta}
\int_0^\infty ds s^{k-2\over2}e^{-s(M^2-\tmu^2)}
\sum_{n\in{\bf Z},\not=0}e^{-s({2n\pi\over\beta})^2}
\mbox{cos}(4n\pi s\tmu/\beta)\ .
\eeq
The summation can be converted to the integrals on the contour 
$C_\mu$ (see Fig.1a):
\beq{intcr}
{\cal O}_\beta^{(k)}(M)={1\over2}(4\pi)^{1\over2}
\int_0^\infty ds s^{k-2\over2}e^{-sM^2}{1\over2\pi i}
\int_{C_\mu}{e^{sz^2}\over e^{\beta(z-\tmu)}-1}dz 
+(\tmu\ra-\tmu)\ .
\eeq
After performing the $s$-integration, one may 
deform the contour $C_\mu$ into another one $C_\delta$ (Fig.1b) 
in the same way as Ref.\cite{HS}, thus obtaining
\beq{OO}
{\cal O}_\beta^{(k)}(M)= \sqrt{\pi}\Gamma({k\over2})
{1\over2\pi i} \int_{\infty}^{(M+)}  
{(M^2-z^2)^{-k/2}\over e^{\beta(z-\tmu)}-1}dz 
+(\tmu\ra-\tmu)\ .
\eeq
Parametrizing the circular part of $C_\delta$ (centered at $M$) 
by $z=M(1-\delta e^{i\phi})$, and using $p^2=z^2-1$ 
for the remaining parts of the contour, we arrive at 
\beq{secondobeta}
{\cal O}_\beta^{(k)}(M)={2\over\sqrt\pi}\Gamma({k\over2})M^{1-k}
\mbox{sin}({2-k\over2}\pi)\Bigl[ \int_{\sqrt{2\delta}}^\infty
{p^{1-k}(p^2+1)^{-1/2}\over e^{\beta X_+}-1}dp +
{{1\over2-k}(2\delta)^{1-{k\over2}}\over e^{\beta(M-\tmu)}-1}
+(\tmu\ra-\tmu)\Bigr] \ .
\eeq
Substituting $k=3$ and $\eps=2$, Eq.\eq{secondobeta} reproduces 
the result (without background field) of Ref.\cite{HS}. 

%
%
\vspace{5mm}
\begin{minipage}[t]{14.5cm} 
%
%
\font\thinlinefont=cmr5
\begingroup\makeatletter\ifx\SetFigFont\undefined
\def\x#1#2#3#4#5#6#7\relax{\def\x{#1#2#3#4#5#6}}%
\expandafter\x\fmtname xxxxxx\relax \def\y{splain}%
\ifx\x\y   
\gdef\SetFigFont#1#2#3{%
  \ifnum #1<17\tiny\else \ifnum #1<20\small\else
  \ifnum #1<24\normalsize\else \ifnum #1<29\large\else
  \ifnum #1<34\Large\else \ifnum #1<41\LARGE\else
     \huge\fi\fi\fi\fi\fi\fi
  \csname #3\endcsname}%
\else
\gdef\SetFigFont#1#2#3{\begingroup
  \count@#1\relax \ifnum 25<\count@\count@25\fi
  \def\x{\endgroup\@setsize\SetFigFont{#2pt}}%
  \expandafter\x
    \csname \romannumeral\the\count@ pt\expandafter\endcsname
    \csname @\romannumeral\the\count@ pt\endcsname
  \csname #3\endcsname}%
\fi
\fi\endgroup
\mbox{\beginpicture
\setcoordinatesystem units <0.68000cm,0.68000cm>
\unitlength=0.68000cm
\linethickness=1pt
\setplotsymbol ({\makebox(0,0)[l]{\tencirc\symbol{'160}}})
\setshadesymbol ({\thinlinefont .})
\setlinear
%
%
\linethickness=0.500pt
\setplotsymbol ({\thinlinefont .})
\circulararc 158.522 degrees from 17.145 22.225 center at 16.223 21.886
\circulararc 158.522 degrees from 15.240 21.907 center at 16.223 21.929
%
%
\linethickness=0.500pt
\setplotsymbol ({\thinlinefont .})
\ellipticalarc axes ratio  0.794:0.794  360 degrees 
	from  6.509 21.907 center at  5.715 21.907

%
%
\linethickness=0.500pt
\setplotsymbol ({\makebox(0,0)[l]{\tencirc\symbol{'170}}})
\putrule from  5.715 25.559 to  5.715 17.304
%
%
\linethickness=0.500pt
\setplotsymbol ({\makebox(0,0)[l]{\tencirc\symbol{'170}}})
\putrule from  1.429 20.796 to 10.795 20.796
%
%
\linethickness=0.500pt
\setplotsymbol ({\makebox(0,0)[l]{\tencirc\symbol{'170}}})
\putrule from  9.207 25.559 to  9.207 17.462
%
%
\linethickness=0.750pt
\setplotsymbol ({\thinlinefont .})
\putrule from  9.207 22.225 to  9.207 22.384
\plot  9.322 22.000  9.207 22.384  9.092 22.000 /
%
%
\linethickness=0.750pt
\setplotsymbol ({\thinlinefont .})
\putrule from  6.509 21.907 to  6.509 21.749
\plot  6.394 22.133  6.509 21.749  6.624 22.133 /
%
%
\linethickness=0.500pt
\setplotsymbol ({\makebox(0,0)[l]{\tencirc\symbol{'170}}})
\putrule from  2.699 25.559 to  2.699 17.304
%
%
\linethickness=0.75pt
\setplotsymbol ({\thinlinefont .})
\putrule from  2.699 22.225 to  2.699 22.066
\plot  2.584 22.450  2.699 22.066  2.814 22.450 /
%
%
\linethickness=0.500pt
\setplotsymbol ({\makebox(0,0)[l]{\tencirc\symbol{'170}}})
\putrule from 17.145 22.225 to 22.860 22.225
%
%
\linethickness=0.500pt
\setplotsymbol ({\makebox(0,0)[l]{\tencirc\symbol{'170}}})
\putrule from 17.145 21.590 to 22.860 21.590
%
%
\linethickness=0.75pt
\setplotsymbol ({\thinlinefont .})
\putrule from 19.844 22.225 to 20.003 22.225
\plot 19.500 22.110 20.003 22.225 19.500 22.340 /
%
%
\linethickness=0.75pt
\setplotsymbol ({\thinlinefont .})
\putrule from 19.844 21.590 to 19.685 21.590
\plot 20.185 21.705 19.685 21.590 20.185 21.475 /
%
%
\put{\SetFigFont{8}{9.6}{rm}x} [lB] at  5.609 24.606
\put{\SetFigFont{8}{9.6}{rm}x} [lB] at  5.609 23.178
\put{\SetFigFont{8}{9.6}{rm}x} [lB] at  5.609 21.749
\put{\SetFigFont{8}{9.6}{rm}x} [lB] at  5.609 20.320
\put{\SetFigFont{8}{9.6}{rm}x} [lB] at  5.609 18.891
\put{\SetFigFont{8}{9.6}{rm}x} [lB] at  5.609 17.462
\put{$\tmu$} [lB] at  4.250 21.709
\put{$\mbox{Im}\,z=0$} [lB] at 11.113 20.796
\put{$C_\mu$} [lB] at  9.620 17.421
\put{$\mbox{Re}\,\tmu=\mu$} [lB] at  6.050 17.421
\put{$(a)$} [lB] at  5.550 16.421
\put{$(b)$} [lB] at  18.050 16.721
\put{\SetFigFont{8}{9.6}{rm}x} [lB] at 16.140 21.749
\put{$M$} [lB] at 16.140 20.303
\put{$C_\delta$} [lB] at 19.844 20.479
\linethickness=0pt
\putrectangle corners at  1.403 25.584 and 22.885 17.278
\endpicture}

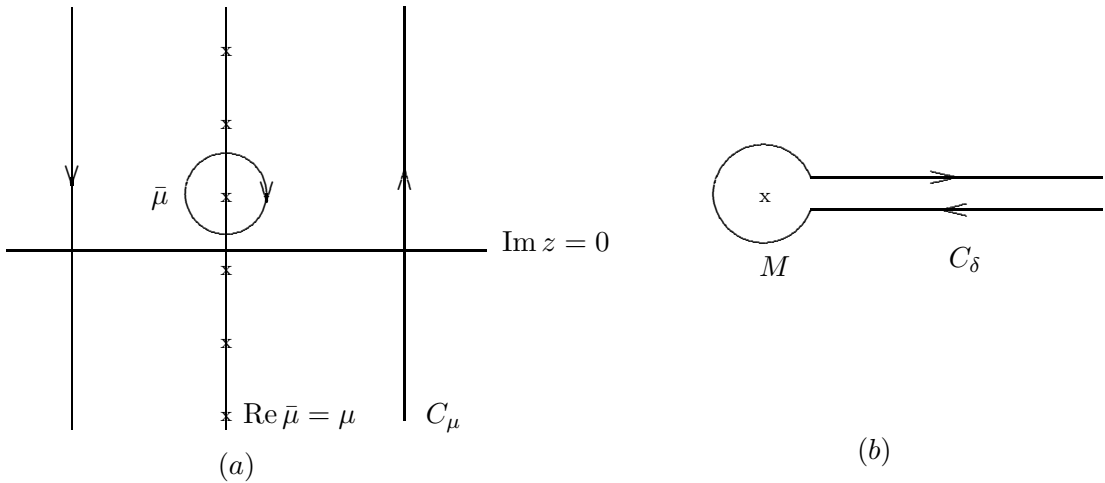
{\bf Figure 1:} The contours $C_\mu$ and $C_\delta$. 
(a) The $C_\mu$ impounds an infinite number of poles at 
$z=\tmu+2n\pi i$, where $n$ denotes nonzero integers. 
(b) Coming from $z=\infty + i\eps$ and going 
to $z=\infty-i\eps$, the $C_\delta$ impounds the 
pole at $z=M$. 
\end{minipage}
\vspace{10mm}

\noindent
Here are a few remarks on the computational difference between 
Ref.\cite{HS} and ours. We have extracted the pure thermodynamic 
part~\eq{pureO} from the beginning, and this fact is expressed 
by the circle at $\tmu$ in $C_\mu$. On the other hand in 
the method of Ref.\cite{HS}, one needs an algebra to separate 
the non-thermodynamic part by using 
\beq{algebra}
{-1\over e^{\beta(z-\tmu)}-1}= 1+
{1\over e^{-\beta(z-\tmu)}-1}\ .
\eeq
The first term on the r.h.s. of Eq.\eq{algebra} corresponds to 
the non-thermodynamic part, and the second term to the following 
expression instead of Eq.\eq{intcr}:
\beq{HSform}
{\cal O}^{(k)}_{\beta\mu}(M)= (4\pi)^{1\over2}
\int_0^\infty ds s^{k-2\over2}
e^{-sM^2}{1\over2\pi i}\Bigl\{
\int_{C_+}{e^{s(z-i{\bar\omega})^2}\over 
e^{\beta(z-\tmu-i{\bar\omega})}-1}  +
\int_{C_-}{e^{s(z-i{\bar\omega})^2}\over 
e^{-\beta(z-\tmu-i{\bar\omega})}-1} \Bigr\}dz \ ,
\eeq
where the contours $C_\pm$ run from $\mu\pm o-i\infty$ to 
$\mu\pm o+i\infty$ for a small value of $o$. Although 
one can of course obtain the same result \eq{secondobeta} 
from Eq.\eq{HSform}, it is not obvious at first glance 
that the $C_-$ integral term corresponds to the 
$\tmu\ra-\tmu$ term of Eq.\eq{OO}.

Now, further results depend on the cases (i) and (ii). 
In the case (i), $k<2$, the second term in the square brackets 
on the r.h.s. of Eq.\eq{secondobeta} is proportional to 
$\delta^{1-k/2}$, and vanishes as $\delta\ra0$. Therefore 
Eq.\eq{secondobeta} reproduces the first representation 
result \eq{o1stcase1}.

In the case (ii), $k\geq2$, the $\delta^{1-k/2}$ term diverges 
as $\delta\ra0$. We expect that it can be canceled with 
the lower surface term from the first integral in the 
square brackets on the r.h.s. of Eq.\eq{secondobeta}. 
This can be seen by a partial integral with applying 
the formula 
\beq{*}
{d\over dp}({\sqrt{p^2+1}\over p})={1\over p^2\sqrt{p^2+1}}
\eeq   
to the $p$-integration in Eq.\eq{secondobeta} as follows:
\beqa
\int_{\sqrt{2\delta}}^\infty
{p^{1-k}(p^2+1)^{-1/2}\over e^{\beta X_+}-1}dp &=&
{{1\over k-2}(2\delta)^{1-k/2}\over e^{\beta(M-\tmu)}-1}
\label{cancel} \\
&+&{1\over2-k}\int_0^\infty p^{3-k}
\Bigl\{ {(p^2+1)^{-3/2}\over e^{\beta X_+}-1} +
{M\beta(p^2+1)^{-1}e^{\beta X_+}\over(e^{\beta X_+}-1)^2}\Bigr\}
dp \nn\ .
\eeqa
The first term on the r.h.s. of Eq.\eq{cancel} cancels the 
divergence as expected, and we derive
\beqa
{\cal O}_\beta^{(k)}(M)&=&
\sqrt{4\pi}\Gamma({k\over2})M^{1-k}
{\mbox{sin}({2-k\over2}\pi)\over(2-k)\pi} \label{semifinal}\\
&\times&\Bigl[\int_0^\infty p^{3-k}
\Bigl\{ {(p^2+1)^{-3/2}\over e^{\beta X_+}-1} +
{M\beta(p^2+1)^{-1}e^{\beta X_+}\over(e^{\beta X_+}-1)^2}\Bigr\}
dp +(\tmu\ra-\tmu)\Bigr]\ . \nn
\eeqa
Putting $t^2=p^2+1$ in Eq.\eq{semifinal}, we finally obtain 
the following concise expression: 
\beq{o2ndcase2}
{\cal O}_\beta^{(k)}(M)=\sqrt{4\pi}\Gamma({k\over2})M^{1-k}
{\mbox{sin}({2-k\over2}\pi)\over(2-k)\pi}\int_1^\infty 
\sqrt{t^2-1}^{2-k}{\partial\over\partial t}\Bigl[{t^{-1}
\over1-e^{\beta(Mt+\tmu)}}\Bigr]dt+(\tmu\ra-\tmu)\ ,
\eeq
which is an alternative representation of Eq.\eq{o1stcase2}. 
As a result, we have derived two expressions in each case: 
Eqs.\eq{basicform} and \eq{o1stcase1} in the case (i), and 
Eqs.\eq{o1stcase2} and \eq{o2ndcase2} in the case (ii). 
In Eq.\eq{o1stcase2} we have to perform $k-2$ derivatives 
(after one integration), while just one integration in 
Eq.\eq{o2ndcase2}, whose integrand thus contains a common 
Boltzmann factor for all $k$ ($\geq2$).   

Let us check the consistency of 
Eq.\eq{o2ndcase2} with the previous result \eq{o1stcase2} for 
$k=2,3,4$. The $k=2$ case \eq{obk2} is immediate from the 
representation \eq{o2ndcase2}, however let us handle these three 
cases simultaneously. Using $(p^2+1)^{-1}=1-p^2(p^2+1)^{-1}$, we 
first divide the second term in the curly brackets in 
Eq.\eq{semifinal}, and then perform a partial integral on 
the r.h.s. in the following quantity:
\beq{qv}
-\int_0^\infty p^{3-k}{M\beta p^2(p^2+1)^{-1}e^{\beta X_+}
\over(e^{\beta X_+}-1)^2}dp=\int_0^\infty p^{3-k}
{p\over\sqrt{p^2+1}}{\partial\over\partial p}
({1\over e^{\beta X_+}-1})dp \ .
\eeq
Changing the integration variable by $t^2=p^2+1$, we arrive at 
\beq{*}
{\cal O}_\beta^{(k)}(M)=
\sqrt{4\pi}\Gamma({k\over2})M^{1-k}
{\mbox{sin}({2-k\over2}\pi)\over(2-k)\pi}
\Bigl[\, I_k + (\tmu\ra-\tmu)\,\Bigr]\ ,
\eeq
where
\beq{*}
I_k=\int_1^\infty\sqrt{t^2-1}^{2-k}t\Bigl[
{k-3\over t}N(t)-  N'(t)\Bigr]dt
\eeq
and $N'(t)\equiv\der_t N(t)$ means the derivative of the function
\beq{*}
N(t)={1\over e^{\beta(Mt-\tmu)}-1}\ .
\eeq
{}~For the $k=2$ and 3 cases, this expression is sufficient to 
see the consistency in each. However, for the $k=4$ case, notice 
that we have to extract a singularity, which cancels a zero from 
the sine function in Eq.\eq{o2ndcase2}, and we hence perform a 
partial integral once more:
\beq{*}
I_k=-\int_1^\infty{1\over4-k}\sqrt{t^2-1}^{4-k}
{\der\over\der t}\Bigl[(k-3){N(t)\over t}-N'(t)\Bigr]\,dt\ ; 
\qquad 2\leq k<4\ ,
\eeq
where the surface terms can be dropped only when $k<4$ (also 
in Eq.\eq{qv}), and this is the reason why this formula is valid 
for $2\leq k<4$ ($-1<d\leq1$). Substituting $k=2,3,4-\epsilon$ 
(with $\eps\ra0$), we verify that these expressions for $I_k$ 
reproduce Eqs.\eq{obk2}-\eq{obk4}. In this sense, 
Eq.\eq{o2ndcase2} is equivalent to Eq.\eq{o1stcase2}. 

{}~For further values of $k$, one should repeat the similar 
calculation for each interval between zero points of the 
sine function. {}~For example, for $k=6$ and 8, we obtain 
\beqa
{\cal O}_\beta^{(6)}(M)&=&{\pi\over16M^5}\Bigl[\,
3N(1)-3N'(1)+N''(1)\,\Bigr]\ , \\
{\cal O}_\beta^{(8)}(M)&=&{\sqrt{\pi}\over8M^7}\Bigl[\,
15N(1)-15N'(1)+6N''(1)-N'''(1)\,\Bigr]\ .
\eeqa
These pure thermodynamic `partial' amplitudes exactly 
correspond to the $N=4$ and 5 pure thermodynamic amplitudes 
in $D=3$ when ${\cal V}_{\beta\mu}={\cal V}$ (the first category). 
In $D=4$, odd $k$ integers are the similar cases. 
{}~For the second category; for example, the $N$-photon amplitude
in $D=4$, we need to combine the quantities from $k=-1$ to $2N-5$. 
These cases can not get rid of non-integrated quantities such 
as Eq.\eq{obk3}. After all, Eqs.\eq{o1stcase2} and \eq{o2ndcase2} 
describe the general amplitude formulae which contain all 
Feynman diagrams. 

\section{The $\beta\ra\infty$ limit}\label{sec6}
\setcounter{section}{6}
\setcounter{equation}{0}
\indent

In the above arguments, we have assumed $\mu<M$, and all 
$\beta\ra\infty$ limits vanish because of it:
\beq{*}
{\cal O}_\beta^{(k)} \quad \alim{\beta}\quad 0\ .
\eeq
To obtain a nontrivial (nonzero) limit, we should remove 
this condition after all. In this sense, the $\mu$-dependence 
of the master amplitudes is non-analytic. To see this, 
we need to transform the function ${\cal O}_\beta^{(k)}$ in a 
form indicating a Bose (or a Fermi) distribution, and we already 
derived this kind of representations in Sections~\ref{sec4} 
and \ref{sec5}. 

Let us begin with the case (i) $k<2$; in particular the case of 
$k=0$ (which is also an odd dimensional case $D=2l+1$ by the way). 
{}~From Eq.\eq{basicform} with changing the variable $E=u+M$, 
we have 
\beq{see}
{\cal O}_\beta^{(0)}(M)=-2\sqrt{\pi}\int_M^\infty{dE\over1\pm
e^{\beta(E-\mu')}}+(\mu'\ra-\mu')\ ,
\eeq
where the plus/minus signs correspond to the Fermi/Bose 
statistics, and we put
\beq{*}
\mu'= \mu-i{\bar\omega}\ .
\eeq
Eq.\eq{see} can be interpreted an effective action or total 
energy density with the chemical potential $\mu'$ and the mass 
$M$ ( $<\mu',\, E$ ). Now, we understand in Eq.\eq{see} that 
the ${\cal O}_\beta^{(0)}$ becomes zero as $\beta\ra\infty$ 
if $\mu'<E$, while the ${\cal O}_\beta^{(0)}$ takes a nonzero 
value if $E<\mu'$. The similar arguments apply to the case of 
generic $k$ value in the following way. Since we observe   
\beq{*}
e^{\beta(M-\tmu+u)}=e^{\beta(u+M-\mu)}e^{i\beta({\bar\omega}+
{2\pi\over\epsilon\beta})}\quad\alim{\beta}\quad 0 
\qquad\mbox{for}\quad u<\mu-M\ ,
\eeq
we have the finite upper boundary on the integral \eq{basicform} 
at $u=\mu-M$, thus obtaining 
\beqa
{\cal O}_\infty^{(k)}(M)&\equiv&\lim_{\beta\ra\infty}
{\cal O}_\beta^{(k)}(M)={-2\sqrt\pi\over\Gamma({2-k\over2})}
\int_0^{\mu-M}(u^2+2Mu)^{-k/2}du\,\theta(\mu-M) \\
&=&{-2\sqrt\pi\over\Gamma({4-k\over2})}(2M)^{-k/2}
(\mu-M)^{2-k\over2}F({k\over2},{2-k\over2};{4-k\over2};
{M-\mu\over2M})\,\theta(\mu-M) \ .\nn
\eeqa
The explicit results for $k=-1,0,1$ are 
\beqa
{\cal O}_\infty^{(1)}(M) &=& 
-2\mbox{arccosh}(\mu/M)\,\theta(\mu-M)\ ,\\
{\cal O}_\infty^{(0)}(M)&=&-2\sqrt{\pi}(\mu-M)\,\theta(\mu-M)\ ,
\label{dirac}\\
{\cal O}_\infty^{(-1)}(M)&=&\Bigl[-2\mu\sqrt{\mu^2-M^2}
+2M^2\mbox{arccosh}(\mu/M)\Bigr]\,\theta(\mu-M) \ .
\eeqa
In one of the $k=0$ cases ($D=3$, $N=0$, $l=1$, 
$\eps=2$)~\cite{KST}, Eq.\eq{dirac} coincides with an exact 
result in~\cite{3d}. 

The case (ii) $k\geq2$ can also be estimated in the same way. 
Applying the similar treatment as above to the Boltzmann factor 
in Eq.\eq{o2ndcase2}, we then obtain the nonzero limit given by 
\beq{limOres}
{\cal O}_\infty^{(k)}(M)=-\sqrt{4\pi}\Gamma({k\over2})M^{1-k}
{\mbox{sin}({2-k\over2}\pi)\over(2-k)\pi}\int_1^{\mu/M}
(t^2-1)^{2-k\over2} {dt\over t^2}\,\theta(\mu-M)\ .
\eeq
Performing the $t$-integration, some of explicit results 
(the $k=2,3,4$ cases) can be shown as follows: 
\beqa
{\cal O}_\infty^{(2)}(M)&=& {\sqrt\pi\over M}({M\over\mu}-1)
\,\theta(\mu-M) \ ,\\ 
{\cal O}_\infty^{(3)}(M)&=&- {1\over\mu M}\sqrt{({\mu\over M})^2-1}
\, \theta(\mu-M)\ ,\\
{\cal O}_\infty^{(4)}(M)&=&- {\sqrt\pi\over2 M^3}
\,\theta(\mu-M)\ .
\eeqa
The calculations are rather straightforward for further $k$ 
values, and the general formula \eq{limOres} suffices. 
The $\epsilon$ dependence is gone away from the 
${\cal O}_\infty^{(k)}$. This is natural because there is no 
difference in the kinematical factors (hence in `partial' 
amplitudes) between boson and 
fermion loops at zero temperature. 

\section{Conclusions}\label{sec7}
\setcounter{section}{7}
\setcounter{equation}{0}
\indent

In this paper, we studied the thermodynamic generalization of the 
one-loop amplitudes which can be cast into the BK master formula 
\eq{gammaN} at zero temperature (with zero chemical potential). 
We followed the two-step procedure: first, calculating the path 
integral $S^1$ summation~\eq{path} to introduce the temperature, 
and then performing the shift manipulation \eq{step2} to insert 
the chemical potential in a loop. This procedure has been applied 
parts by parts to the kinematical factor ${\cal K}$, the 
normalization factor ${\cal N}$ and the vertex structure 
function (reduced kinematical factor) ${\cal V}$, and we have 
derived the general formulae for these parts. {}~From 
Eqs.\eq{GBM}, \eq{Vbmexp} and \eq{AN}, the thermodynamic 
$N$-point amplitude of general form is thus summarized as 
\beq{con1}
\Gamma_N^{\beta\mu}=c\sum_{l,n\in {\bf Z}}a_{ln}
{\der^n\over\der\Omega^n}
\int_0^1d\sigma_1\int_0^1d\sigma_2\cdots
\int_0^1d\sigma_N  {\cal A}_l^{\beta\mu} \ .
\eeq 
One can check the validity of this master formula in various cases; 
the free effective potentials and the photon polarization in 
Appendix B, the effective potential in a constant magnetic field 
in Ref.\cite{KST}, and the $\pi^0\ra2\gamma$ decay in 
Ref.\cite{HS} etc. 

The detail analyses on the building blocks ${\cal V}_{\beta\mu}$ 
and ${\cal K}_{\beta\mu}$ have given us useful information. 
We have realized that the total thermodynamic kinematical 
factor ${\cal K}_{\beta\mu}$ behaves as the thermodynamically 
generalized normalization to the zero-temperature type 
correlator shown in Eq.\eq{normb}. This is certainly a useful 
result and makes calculations for large $N$ much simpler than 
ever; namely we have only to attach the new normalization 
${\cal N}_{\beta\mu}$ to the $N$-point scalar correlator 
(with switching $k^0_j\ra\omega_{k_j}$). Another interesting 
point is that the pure thermodynamic part of a vertex structure 
function can be expressed in terms of either vanishing 
or $\Omega$-differential operators, and this fact makes us 
possible to decouple the master formula~\eq{GBM} into 
$\Gamma_N^{\beta\mu}= \Gamma_N +{\tilde\Gamma}_N^{\beta\mu}$ 
in a nontrivial way (cf. Eqs.\eq{GBM} and \eq{GMBt}). 
It is worth noticing that we do not decouple the vertex 
structure function but the kinematical factor only. 
{}~From Eqs.\eq{GMBt}, \eq{Vbmexp} and \eq{AN2}, we then 
conclude that the pure thermodynamic master amplitude is 
simply given by 
\beq{con2} 
{\tilde\Gamma}_N^{\beta\mu}=c\sum_{l,n\in {\bf Z}}a_{ln}
{\der^n\over\der\Omega^n}
\int_0^1d\sigma_1\int_0^1d\sigma_2\cdots
\int_0^1d\sigma_N {\tilde{\cal A}}_l^{\beta\mu} \ .
\eeq 
Apart from $\Gamma_N$, the pure thermodynamic 
part is renormalization free, and hence Eq.\eq{con2} is 
the final form to apply our integral formulae derived in 
Sections~\ref{sec4}-\ref{sec6}. More explicitly, we have 
(from Eqs.\eq{tmu} and \eq{OA})
\beq{con3}
{\tilde\Gamma}_N^{\beta\mu}=c(4\pi)^{-D\over2}
\sum_{l,n\in {\bf Z}}a_{ln}(-i)^n
{\der^n\over\der\tmu^n}
\int_0^1d\sigma_1\int_0^1d\sigma_2\cdots
\int_0^1d\sigma_N {\cal O}_\beta^{(2l+1-D)}(M)
\Bigr|_{m^2\ra m^2+b_l} \ .
\eeq
and in particular for the $\beta\ra\infty$ limit with $\mu\not=0$ 
\beq{con4}
\lim_{\beta\ra\infty}{\tilde\Gamma}_N^{\beta\mu}=
c(4\pi)^{-D\over2}
\sum_{l\in {\bf Z}}a_{l0}
\int_0^1d\sigma_1\int_0^1d\sigma_2\cdots
\int_0^1d\sigma_N {\cal O}_\infty^{(2l+1-D)}(M)
\Bigr|_{m^2\ra m^2+b_l} \ .
\eeq
The thermodynamic master amplitudes \eq{con3} and \eq{con4} are 
provided with Eqs.\eq{o1stcase2}, \eq{o2ndcase2} and \eq{limOres}, 
and several explicit 'partial amplitudes' 
${\cal O}_\beta^{(k)}$ are also given up to $k=8$. 

In the master formulae~\eq{con2} and \eq{con3}, the `partial' 
amplitudes ${\tilde{\cal A}}_l^{\beta\mu}$ or 
${\cal O}_\beta^{(k)}$ ($s$-integrals of the pure thermodynamic 
kinematical factor) are the fundamental computational blocks in 
our formalism, and in Sections~\ref{sec4} and \ref{sec5}, we 
focused on some mathematical aspects and techniques how to 
calculate these integrals. We derived various representations 
and formulae on these integrals for arbitrary values of $N$, 
$D$ and $\epsilon$. It is obvious that a variety of equivalent 
formulae makes easy to find some relation and consistency with 
the results obtained by different methods. These should 
generally be related by analytic continuation, and a different 
analytic expression is certainly useful by definition; i.e., 
another formula can cover the region which can not be reached 
in an original representation, and it also overcomes some 
defect of an inconvenient representation. For example in 
Ref.~\cite{KST}, a dual rotation is used to obtain an electric 
gap equation from magnetic one in a four-fermion model. We 
expect such kind of utility when our formulae are 
applied in more explicit stages. 
 
Although we could have possibly simplified our results 
furthermore through a certain technique \cite{MR}, our formulas 
were sufficiently convenient to extract the nonzero values of 
zero temperature limit with $\mu$ kept finite. {}~For this 
purpose, it is necessary to have the integral representation, 
such as Eqs.\eq{basicform} and \eq{o2ndcase2}, which clearly 
indicates a non-analytic cut in its integrand at 
$\beta=\infty$. We also had to go beyond the condition $\mu<M$ 
in these representations, however this might be justified by 
analytic continuation, and at least we know that the $k=0$ 
case coincides with the exact result. 

We have examined the model independent structure of the 
thermodynamic BK master formula with consulting several simple 
examples. The model dependence appears in the vertex structure 
functions, and hence one has to evaluate the vertex structure in 
the first place for explicit calculations in each model. This 
task is case by case and will become more lengthy as $N$ 
increasing for the cases involving more bosonic $x$-fields in 
the vertex operator $v_j$ parts such as photon, gluon and 
pseudo-scalar particle. However, our systematic prescription 
is certainly promising to obtain the (pure) thermodynamic 
$N$-point amplitudes in a straightforward way as long as the 
${\cal V}_{\beta\mu}$ belongs to the general form \eq{Vbmexp}. 
Finally, we should not forget the advantage that the world-line 
formula encapsulates all necessary Feynman diagrams in a 
single integrand. 

%
%
%

\appendix

\section*{Appendix A. Insertion of chemical potential}
\setcounter{section}{1}
\setcounter{equation}{0}
\indent

In this appendix, we give a brief explanation of the shift 
procedure \eq{shift}. {}~First consider the vacuum case 
${\bar\omega}=0$. In this case, $M^2=m^2$ and 
\beq{Trln}
-\mbox{Trln}(\der^2+m^2)=
{1\over\beta}\sum_n\int{d^Dp\over(2\pi)^{D-1}}
\int_0^\infty{ds\over s}e^{-s(p^2+m^2)} \ .
\eeq
We usually perform the insertion of chemical potential 
in terms of the shift 
\beq{*}
p^2 = p_0^2 + \vec{p}{}^2 \quad\ra\quad 
(\omega_n+i\mu)^2  +  \vec{p}{}^2 
\eeq
with the internal Matsubara frequencies
\beq{*}
\omega_n={2\pi\over\beta}(n+{1\over\epsilon})\ , 
\eeq
and we have
\beq{*}
\sum_n e^{-s(\omega_n+i\mu)^2} = e^{s\tmu_0^2} \Theta_3(
{2s\tmu_0\over\beta},is{4\pi\over\beta^2})\ ,
\eeq
where we have defined 
\beq{*}
\tmu_0 =\mu-{2\pi i\over\eps\beta}\ .
\eeq
Thus, we prove Eq.\eq{secondrep} for $N=0$:
\beq{A0}
\mbox{Eq.\eq{Trln}}=
{1\over\beta}\int_0^\infty{ds\over s}(4\pi s)^{1-D\over2}
e^{s(\tmu_0^2-m^2)}
\Theta_3({2s\tmu_0\over\beta},is{4\pi\over\beta^2})\ .
\eeq
{}~For further nonzero values of $N$, the proofs are 
straightforward, and we shall not explain the details anymore. 
{}~For example, one can find the $N=2$ case in Appendix B.

Instead, we add a supplemental interpretation. Applying the 
transformation
\beq{*}
{\sqrt{4\pi s}\over\beta}e^{s\tmu_0^2}
\Theta_3({2s\tmu_0\over\beta},is{4\pi\over\beta^2}) =
\Theta_3({i\beta\tmu_0\over2\pi},i{\beta^2\over4\pi s})\ ,
\eeq
we rewrite Eq.\eq{A0} as
\beq{*}
\mbox{Eq.\eq{A0}}=
{\cal A}_0^{\beta\mu} =\int_0^\infty{ds\over s}(4\pi s)^{-D\over2}
e^{-sm^2}\Theta_3(i{\beta\tmu_0\over2\pi},i{\beta^2\over4\pi s})\ .
\eeq
Here $\tmu_0$ now appears only in the first argument of the 
$\Theta_3$, and remember that ${\bar\omega}$ also appears in 
the same place as $\tmu_0$ does for $N\not=0$. Taking account of 
exponent's additivity in the first argument of $\Theta_3$ 
( {\it q.v.} Eq.\eq{def} ), one can imagine that the final 
result is given by the replacement of $\tmu_0$ with 
$\tmu_0-i{\bar\omega}$ ( $=\tmu$ ). This result 
coincides with the ${\cal K}_{\beta\mu}$ obtained by the 
shift manipulation \eq{shift} in ${\cal K}_\beta$.  

\section*{Appendix B. Photon self-energy part from Feynman rule}
\setcounter{section}{2}
\setcounter{equation}{0}
\indent

In this appendix, we rearrange the $N=2$ Feynman amplitude of 
photon scattering into the world-line formula at finite $\beta$ 
and $\mu$. The photon self-energy part (in $D=4$) by the Feynman 
diagram technique is written in the form
\beq{*}
\Pi_\beta=-{1\over\beta}\sum_n\int{d^3p\over(2\pi)^3}\,
{\mbox{Tr}[\epslash_1(\pslash-\kslash_1)\epslash_2\pslash\,]
\over p^2(p-k)^2} \ ,
\eeq
where $p^\mu=(\omega_n,\vec{p})$, 
$k^\mu_i=(\omega_{k_i},\vec{k}_i)$, 
and $\eps_i\equiv \eps^i_\mu$, $i=1,2$. We also make use of 
$k^\mu=k^\mu_1=-k^\mu_2$. The QED coupling $e$ is set to be the 
unity for simplicity. {}~For convenience, we decompose the 
$\Pi_\beta$ into the following three parts:
\beq{*}
\Pi_i = {1\over\beta}\sum_n\int{d^3p\over(2\pi)^3}\,W_i\ ,
\eeq
in terms of 
\beq{*}
-\mbox{Tr}[\epslash_1(\pslash-\kslash)\epslash_2\pslash\,]
=W_1+W_2+W_3\ ,
\eeq
where
\beqa
W_1&=& 2\eps_1\cdot\eps_2(p^2+(p-k)^2)\ , \\
W_2&=& 2(\eps_1\cdot\eps_2k_1\cdot k_2
       -\eps_1\cdot k_2\eps_2\cdot k_1)\ ,\\
W_3&=& -2\eps_1\cdot(2p-k_1)\eps_2\cdot(2p-k_1)\ .
\eeqa
Applying the Feynman integral formula 
\beq{feynman}
{1\over p^2(p-k)^2}=\int_0^\infty dss\int_0^1 du e^{-sk^2(u-u^2)}
e^{-s(p-(1-u)k)^2}\ ,
\eeq
we easily rewrite $\Pi_1$ and $\Pi_2$ in the following forms: 
\beq{Pi1}
\Pi_1 = {4\eps_1\cdot\eps_2\over\beta}
\sum_n\int_0^\infty {dss\over(4\pi s)^{3/2}}\int_0^1 du 
e^{-sk^2(u-u^2)}e^{-s(\omega_n+a)^2} 
\,{1\over s}\delta(1-u)\ ,
\eeq
\beq{Pi2}
\Pi_2=
(\eps_1\cdot\eps_2k_1\cdot k_2-\eps_1\cdot k_2\eps_2\cdot k_1)
{2\over\beta}\sum_n
\int_0^\infty {dss\over(4\pi s)^{3/2}}\int_0^1 du 
e^{-sk^2(u-u^2)}e^{-s(\omega_n+a)^2}\ ,
\eeq
where 
\beq{aome}
a=(1-u)\omega_k\ .
\eeq
The $\Pi_3$ can be rewritten in the similar way by using 
Eq.\eq{feynman} and shifting $\vec{p}\ra\vec{p}+(1-u)\vec{k}$ 
in the $p$-integral. After some algebra, we have 
\beqa
\Pi_3&=&
{2\over\beta}\sum\int_0^\infty {ds\over(4\pi s)^{3/2}}s \int_0^1 du
e^{-sk^2(u-u^2)}e^{-s(\omega_n+a)^2}\label{Pi3}\\
&\times&\Bigl[\,{2\over s}(\eps_0^1\eps_0^2-\eps_1\cdot\eps_2)
-\Bigl\{\,
2\eps^1_0(\omega_n+a)-(1-2u)\eps_1\cdot k\,\Bigr\}
\Bigr\{\,2\eps^2_0(\omega_n+a)-(1-2u)\eps_2\cdot k\,\Bigr\}
\,\Bigr]\ .\nn
\eeqa
Gathering Eqs.\eq{Pi1}, \eq{Pi2} and \eq{Pi3}, we obtain 
\beqa
\Pi_\beta&=&{2\over\beta}\sum_n\int_0^\infty 
{dss\over(4\pi s)^{3/2}}\int_0^1 du
e^{-sk^2(u-u^2)}e^{-s(\omega_n+a)^2}\nn\\
&\times&\Bigl[\,{2\over s}\eps_1\cdot\eps_2\{\,\delta(1-u)-1\,\} +
(\eps_1\cdot\eps_2k_1\cdot k_2-\eps_1\cdot k_2\eps_2\cdot k_1)\nn\\
&-&(1-2u)^2\eps_1\cdot k\eps_2\cdot k
-4\eps_0^1\eps_0^2\,\{(\omega_n+a)^2 - {1\over2s}\,\}\nn\\
&+&2(1-2u)(\eps_0^1\eps_2\cdot k+\eps_0^2\eps_1\cdot k)
(\omega_n+a)\,\Bigr]\ .\label{Pibeta}
\eeqa

Now, we can eliminate the forms $(\omega_n+a)^m$; $m=1,2$, 
in the summand of Eq.\eq{Pibeta} due to the following operations:
\beqa
-{1\over2s}\der_a\sum_n\,e^{-s(\omega_n+a)^2}&=&
\sum_n\, (\omega_n+a)e^{-s(\omega_n+a)^2} \ ,\\
({1\over2s}\der_a)^2\sum_n\,e^{-s(\omega_n+a)^2}&=&
\sum_n\, [(\omega_n+a)^2-{1\over2s}]e^{-s(\omega_n+a)^2}\ .
\eeqa
The $a$-dependence of the summand exponential can be transformed 
into a linear exponent form by using the Jacobi transformation:
\beqa
\sum_n e^{-s(\omega_n+a)^2}&=&
e^{-sa^2} \Theta_2({2sia\over\beta},{is4\pi\over\beta^2})=
{\beta\over\sqrt{4\pi s}}
\Theta_4({\beta a\over2\pi},{i\beta^2\over4\pi s})\nn\\
&=&{\beta\over\sqrt{4\pi s}}\sum_n(-1)^ne^{-{n^2\beta^2\over4s}}
e^{in\beta a}\ . 
\eeqa
Then we are allowed to perform the following replacement 
in the transformed summands 
\beq{*}
{\der\over\der a}\quad \ra\quad in\beta\ ,
\eeq
and we finally derive the formula 
\beqa
\Pi_\beta&=& 2\int_0^\infty{dss\over(4\pi s)^{4/2}}\int_0^1 du
e^{-sk^2(u-u^2)}\sum_n(-1)^n 
e^{-{n^2\beta^2\over4s}}e^{in\beta a}\nn\\
&\times&\Bigl[\,
\eps_0^1\eps_0^2{n^2\beta^2\over s^2} +{in\beta\over s}(2u-1)
(\eps_0^1\eps_2\cdot k_1- \eps_0^2\eps_1\cdot k_2)\nn\\
&+&{2\over s}\eps_1\cdot\eps_2\{\,\delta(1-u)-1\,\} +
(\eps_1\cdot\eps_2k_1\cdot k_2-\eps_1\cdot k_2\eps_2\cdot k_1)\nn\\
&+&(1-2u)^2\eps_1\cdot k_2\eps_2\cdot k_1\,\Bigr]\ .\label{Pibetaf}
\eeqa

{}~For $\mu\not=0$, we have only to make the standard 
shift \eq{matsu}, and this causes the following change of the 
$a$ defined in Eq.\eq{aome}: 
\beq{*}
a    \quad\ra\quad   (1-u)\omega_k + i\mu \ .
\eeq
This modification exactly corresponds to the shift \eq{shift}. 
Note that the $a$ coincides with ${\bar\omega}$ with fixing 
$\sigma_1=1$.

\section*{Appendix C. World-line method for photon self-energy part}
\setcounter{section}{3}
\setcounter{equation}{0}
\indent

In this appendix, we illustrate how we obtain the thermodynamic 
version of the vertex structure function in the case of 
$N=2$ photon scattering. In the world-line formalism, 
the photon self-energy part at zero temperature can be 
obtained from the formula 
\beq{Pi0}
\Gamma_2= 
-{1\over2}2^{D\over2}\int_0^\infty{ds\over s}e^{-sm^2}
\oint{\cal D}x{\cal D}\psi\,\exp[-\int_0^s({1\over4}{\dot x}(\tau)
+{1\over2}\psi{\dot\psi})d\tau\,]
V_1V_2\ ,
\eeq
where $V_j$, $j=1,2$, are the photon vertex operators  
\beq{photon}
V_j=-ie\int_0^s d\tau_j(\eps_j\cdot{\dot x}+2i\psi\cdot\eps_j
\psi\cdot k_j)(\tau_j)e^{ik_j\cdot x(\tau_j)}\ ,
\eeq
and ${\dot x}=\der_\tau x$ etc. Here we follow the standard 
world-line notation $\tau$, which is related to the main text 
notation $\sigma$ by 
\beq{*}
\tau_j=s\sigma_j \ .
\eeq  

Eq.\eq{Pi0} is known to become~\footnote{We set the QED 
coupling $e=1$ in the following (as in Appendix B).}
\beq{Pi0strac}
\Gamma_2=-{1\over2}2^{D\over2}\int_0^\infty{ds\over s}
\int_0^s d\tau_1 \int_0^s d\tau_2\,
{\cal V} \times {\cal K} 
\eeq
with the vertex structure function 
\beq{Vsphoto}
{\cal V}=
\eps_1\cdot\eps_2 {\ddot G}_B^{12} +\eps_1\cdot k_2\eps_2\cdot k_1
({\dot G}_B^{12})^2 +(\eps_1\cdot\eps_2 k_1\cdot k_2 -
\eps_1\cdot k_2\eps_2\cdot k_1)(G_F^{12})^2\ ,
\eeq
and the kinematical factor 
\beqa
{\cal K}&=&{\cal N}<e^{ik_1\cdot x(\tau_1)}e^{ik_2\cdot x(\tau_2)}> 
=e^{-sm^2}\oint{\cal D}x e^{-\int_0^s{1\over4}{\dot x}^2d\tau}
\prod_{j=1}^2 e^{ik_j\cdot x(\tau_j)}\nn\\
&=& (4\pi s)^{-D/2}e^{-sm^2}e^{k_1\cdot k_2 G_B^{12}}\ ,
\eeqa
where
\beq{*}
G_B^{12}=|\tau_1-\tau_2|-{(\tau_1-\tau_2)^2\over s}, 
\qquad G_F^{12}=\mbox{sign}(\tau_1-\tau_2)\ .
\eeq
The path integral normalizations are chosen to be
\beq{*}
\oint{\cal D}x e^{-\int_0^s{1\over4}{\dot x}^2d\tau}=
(4\pi s)^{-D/2}\ , \qquad 
\oint{\cal D}\psi e^{-\int_0^s\psi\cdot{\dot\psi}d\tau}=1\ .
\eeq
It can be said that the kinematical factor is defined by the 
insertions of the $\phi^3$ scalar vertex operators
\beq{*}
V_j=\int_0^s d\tau_j \exp[ik_j\cdot x(\tau_j)]\ ;\qquad j=1,2\ .
\eeq

According to the program presented in Section~\ref{sec3}, 
we are led to calculate the following path integral 
at finite temperature: 
\beqa
\Gamma_2^\beta&=& -{1\over2}2^{D\over2}(ie)^2
\int_0^\infty{ds\over s}e^{-sm^2}
\oint{\cal D}x{\cal D}\psi\,\exp[-\int_0^s({1\over4}{\dot x}^2
+{1\over2}\psi\cdot{\dot\psi})d\tau]    \nn\\
&\times& (\prod_{i=1}^2\int_0^sd\tau_i e^{ik_ix(\tau_i)})
\sum_n(-1)^n e^{-{n^2\beta^2\over4s}}
e^{in{\beta\over s}(\tau_1-\tau_2)\omega_k}\nn\\
&\times&(\eps_0^1{n\beta\over s}+\eps^1\cdot {\dot x}
+2i\psi\cdot\eps_1\psi\cdot k_1)(\tau_1)
(\eps_0^2{n\beta\over s}+\eps^2\cdot {\dot x}
+2i\psi\cdot\eps_2\psi\cdot k_2)(\tau_2)\ .\label{detail}
\eeqa
Using the Wick contraction method with the correlators
\beq{*}
<x^\mu(\tau_1)x^\nu(\tau_2)>=-g^{\mu\nu}G_B^{12}\ ,
\qquad 
<\psi^\mu(\tau_1)\psi^\nu(\tau_2)>={1\over2}g^{\mu\nu}G_F^{12}\ ,
\eeq
one can verify the coincidence of the $\Gamma_2^\beta$ with the 
$\Pi_\beta$ derived in Appendix B; we have arrived at the form
\beq{PB}
\Pi_\beta = {1\over2}2^{D\over2}
\int_0^\infty{ds\over s}e^{-sm^2} (\prod_{j=1}^2\int_0^sd\tau_j)
\sum_n(-1)^n {\cal K}^{(n)}_\beta 
(\, {\cal V} + {\tilde{\cal V}}\,) \ ,
\eeq
where
\beq{nut}
{\tilde{\cal V}} = \eps_0^1\eps_0^2{n^2\beta^2\over s^2} 
+{in\beta\over s}(\eps_0^1\eps_2\cdot k_1-
\eps_0^2\eps_1\cdot k_2){\dot G}_B^{12}\ ,
\eeq
and the ${\cal K}^{(n)}_\beta$ is the $n$th mode of the bosonic 
two point function at finite $\beta$ defined by 
\beqa
{\cal K}^{(n)}_\beta&=&\oint{\cal D}x
e^{-\int_0^s{1\over4}{\dot x}^2} \prod_{j=1}^2 
e^{ik_j\cdot x(\tau_j)}\Bigr|_{x^0\ra x^0+n\beta\tau/s}
\nn\\
&=& e^{-{n^2\beta^2\over4s}}e^{in\beta(\tau_1-\tau_2)\omega_k/s}
(4\pi s)^{-D/2}e^{-k^2G_B^{12}}\ .\label{kbetan}
\eeqa
Here Eq.\eq{PB} with \eq{nut} and \eq{kbetan} reproduces the 
Feynman rule result \eq{Pibetaf} with rescaling $\tau_i=s\sigma_i$ 
and fixing $\sigma_1=1$ with $\sigma_2=u$. Note that 
${\dot G}_B^{12}$ behaves as $2u-1$ in this respect. 

Let us further rewrite the above ${\tilde{\cal V}}$ in an 
operator form suitable to Eq.\eq{GBM}. In the following, we shall 
not employ the fixing $\sigma_1=1$. As shown in Appendix B, we 
can replace $in\beta$ with $\der/\der{\bar\omega}$, where 
\beq{ome12}
{\bar\omega}=(\sigma_1-\sigma_2)\omega_k \ ,
\eeq
and we therefore find
\beq{easyPi}
\Gamma_2^\beta={1\over2}2^{D\over2}\int_0^\infty ds s 
\,(\prod_{i=1}^2\int_0^1d\sigma_i)\,
{\cal V}_\beta \times {\cal K}_\beta 
\eeq
with having
\beq{Vsbphoto}
{\cal V}_\beta = {\cal V} 
-\eps_0^1\eps_0^2({1\over s}{\der\over\der{\bar\omega}})^2 
-{1\over s}{\der\over\der{\bar\omega}}(\eps_0^1\eps_2\cdot k_1-
\eps_0^2\eps_1\cdot k_2){\dot G}_B^{12}
\eeq
and 
\beq{*}
{\cal K}_\beta = \sum_n(-1)^n {\cal K}_\beta^{(n)} e^{-sm^2}\ ,
\eeq
where the change of variables $\tau_i=s\sigma_i$ in $G_B^{12}$ 
should be understood. Since we already justified the 
shift~\eq{shift} in the end of Appendix B, we can use the 
following parts for the finite $\mu$ case   
\beq{Vsbmphoto}
{\cal V}_{\beta\mu}={\cal V}_{\beta}
[\,{\der\over\der{\bar\omega}}\,\ra\, 
{\der\over\der\Omega}\,]\ ,\qquad 
{\cal K}_{\beta\mu}={\cal K}_\beta[\,{\bar\omega}\,\ra\,
\Omega\,]\ .
\eeq
Substituting these in Eq.\eq{easyPi}, we obtain the two-point 
function $\Gamma_2^{\beta\mu}$, which therefore coincides with 
the corresponding Feynman rule result suggested in the end of 
Appendix B.


\end{document}